\documentclass[a4paper,10pt]{article}
\pdfoutput=1

\usepackage{jcappub}

\usepackage[style=base, singlelinecheck=off, font=small, labelfont=bf, 
justification=raggedright]{caption}

\usepackage{subfigure}

\usepackage[utf8]{inputenc}

\usepackage{lipsum}
\usepackage{mwe}

\usepackage{hyperref}
\hypersetup{
    colorlinks=true,
    linkcolor=blue,
    filecolor=magenta,      
    urlcolor=blue,
}

\usepackage[bottom]{footmisc}
\usepackage{float}
\usepackage{bm}
\usepackage{latexsym}
\usepackage{graphicx,wrapfig}
\usepackage[export]{adjustbox}
\usepackage{color}
\usepackage[dvipsnames]{xcolor}
\usepackage{verbatim}
\usepackage{array}
\usepackage{todonotes}
\usepackage{gensymb}
\usepackage{amsmath}

\def\Eq#1{eq.~\eqref{#1}}
\newcommand{\bs}[1]{\boldsymbol{#1}}

\newcommand{\calE}{\mathcal{E}}
\newcommand{\calH}{\mathcal{H}}

\newcommand{\calP}{\mathcal{P}}
\newcommand{\pd}{\partial}

\title{Bayesian estimation of our local motion from the Planck-2018 CMB temperature map}

\author[1]{Sayan Saha,} 
\emailAdd{sayan.saha@students.iiserpune.ac.in} 
\affiliation [1]{Department of Physics, Indian Institute of Science Education and Research, Pune 411008, India}

\author[2, 3]{Shabbir Shaikh,}
\emailAdd{sshaik14@asu.edu } 
\affiliation [2]{School of Physical Sciences, National Institute of Science Education and Research, HBNI, Jatni 752050,
Odisha, India}
\affiliation[3]{School of Earth and Space Exploration, Arizona State University, Tempe, AZ 85287, USA}

\author[4,5,6,7]{Suvodip Mukherjee,} 
\emailAdd{smukherjee1@perimeterinstitute.ca} 
\affiliation [4]{Gravitation Astroparticle Physics Amsterdam (GRAPPA), Anton Pannekoek Institute for Astronomy and Institute for Physics, 
University of Amsterdam, Science Park 904, 1090 GL Amsterdam, The Netherlands}
\affiliation [5]{Institute Lorentz, Leiden University, PO Box 9506, Leiden 2300 RA, The Netherlands}
\affiliation [6]{Delta Institute for Theoretical Physics, Science Park 904, 1090 GL Amsterdam, The Netherlands}
\affiliation [7]{Perimeter Institute for Theoretical Physics, 31 Caroline Street N., Waterloo, Ontario, N2L 2Y5, Canada}

\author[1,8]{Tarun Souradeep,} 
\emailAdd{tarun@iiserpune.ac.in} 
\affiliation [8]{Inter University Centre for Astronomy and Astrophysics, Post Bag 4, Ganeshkhind, Pune-411007, India}

\author[9,10,11]{Benjamin D. Wandelt}
\emailAdd{bwandelt@iap.fr}
\affiliation[9]{Institut d'Astrophysique de Paris (IAP), UMR 7095, CNRS/UPMC Universit\'e Paris 6, Sorbonne Universit\'es, 98 bis boulevard Arago, F-75014 Paris, France}
\affiliation[10]{Institut Lagrange de Paris (ILP), Sorbonne Universit\'es, 98 bis Boulevard Arago, 75014 Paris, France}
\affiliation[11]{Center for Computational Astrophysics, Flatiron Institute, 162 5th Avenue, New York, NY 10010, USA}

\date{\today}
\abstract{
 The largest fluctuation in the CMB sky is the CMB dipole, which is believed to be caused by the motion of our observation frame with respect to the CMB rest frame. This motion accounts for the known motion of the Solar System barycentre with a best-fit amplitude of 369 km/s, in the direction ($\ell= 264^\circ$, $b=48^\circ$) in galactic coordinates. Along with the CMB dipole signal, this motion also causes an inevitable signature of statistical anisotropy in the higher multipoles due to the modulation and aberration of the CMB temperature and polarization fields. This leads to a correlation between adjacent CMB multipoles causing a non-zero value of the off-diagonal terms in the covariance matrix which can be captured in terms of the dipolar spectra of the bipolar spherical harmonics (BipoSH). In our work, we jointly infer the CMB power spectrum and the BipoSH spectrum in a Bayesian framework using the \textit{Planck}-2018 \texttt{SMICA} temperature map. We detect amplitude and direction of the local motion consistent with the canonical value $v=369$ km/s  inferred from CMB dipole with a statistical significance of $4.54\sigma$, $4.97\sigma$ and $5.23\sigma$ respectively from the masked temperature map with the available sky fraction $40.1\%$, $59.1\%$, and $72.2\%$, confirming the common origin of both the signals. The Bayes factor in favor of the canonical value is between $7$ to $8$ depending on the choice of mask. But it strongly disagrees (by a value of the Bayes factor about $10^{-10}-10^{-11}$) with a higher value of local motion which one can infer from the amplitude of the  dipole signal obtained from the CatWISE2020 quasar catalog using the WISE and NEOWISE data set.}

\begin{document}

\maketitle

\section{Introduction}\label{Introduction}

The relative motion of the Solar System with respect to the rest frame of the Cosmic Microwave Background (CMB) modifies the statistical properties of the CMB temperature and polarization anisotropies. It is known to be the major contributor to the dipolar $\ell=1$ fluctuation in the observed CMB anisotropy, the biggest known fluctuation in the observed CMB, which causes mK ($10^{-3}$ K) fluctuations on the mean CMB temperature $T_0 = (2.7255 \pm 0.0006) K$ \cite{1993ApJ...419....1K,Fixsen_2009}. The velocity of the Solar System barycentre, inferred from the CMB dipole measurement, has been found to be $v=$ (369$\pm$0.9) km/s in the direction $(\ell,b)$ = (263.99$\pm$0.14$\degree$, 48.26$\pm$0.03$\degree$) \cite{1993ApJ...419....1K,1994ApJ...420..445F, Lineweaver:1996xa, 1996ApJ...473..576F, Hinshaw_2009}. 

\begin{figure}
\centering
\includegraphics[width=0.8\textwidth]{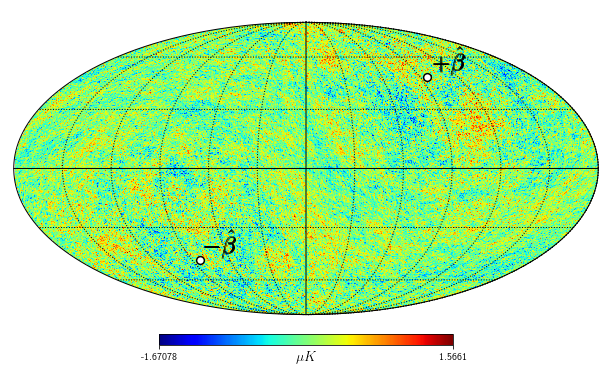}
\caption{We show the difference between a simulated statistically isotropic map and a Doppler boosted map at 217 GHz generated with the same value of the initial random seed using \texttt{CoNIGS} \cite{Mukherjee:2013kga} with an injected signal in the fiducial dipole direction $\ell=  264^\circ$, $b= 48^\circ$, with the known amplitude $v/c \equiv \beta= 1.23\times10^{-3}$ for the illustration purpose.}\label{map_difference}
\end{figure}

Along with the well-known dipole anisotropy, this motion also causes correlation between the neighbouring spherical harmonic coefficients $l$ and $(l\pm1)$ of the observed CMB sky \cite{Challinor:2002zh, Kamionkowski:2002nd, Mukherjee:2013zbi}, which leads to non-zero off-diagonal terms in the harmonic space covariance matrix of temperature and polarization field, which can be captured in terms of the dipole spectra of the bipolar spherical harmonics (BipoSH) \cite{Hajian_2003, Hajian:2005jh, Souradeep_2006}. The Lorentz transformation of the CMB photons in our observation frame gives rise to two effects in the relative intensity fluctuations, \textit{modulation} and \textit{aberration}, which we jointly refer as \textit{Doppler boost}. The first one is a frequency dependent effect that causes the CMB sky to appear brighter (or dimmer) by an amount\footnote{The speed of light in vacuum is denoted by $c=2.997\times 10^5$ km/s.} $\beta \equiv v/c$ = ($1.23\times10^{-3}$) in the direction (or opposite direction) of motion. The second effect, aberration is frequency independent and results in the deflection of the original directions of the incoming CMB photons. The strength of Doppler boost varies across the sky and it depends on the projection ($\hat \beta\cdot\hat n$) of the direction of the incoming  photons ($\hat n$) and the direction of local motion ($\hat \beta$). The imprints of these effects in the CMB temperature field is shown in Figure \ref{map_difference}, where we plot the difference between a simulated Doppler boosted CMB map at 217 GHz (generated using \texttt{CoNIGS} \cite{Mukherjee:2013kga}) with the known value of the velocity amplitude $\beta = 1.23\times10^{-3}$ in the known dipole direction (shown by the black plus symbol) and a statistically isotropic (SI) map generated with the same initial random seed. This difference map shows the fluctuations at all angular scales as a result of the scale-independent nature of Doppler boost. The imprints of Doppler boost vary with the sky direction by the factor $\hat \beta\cdot\hat n$. The effect is stronger in the direction (or in the opposite direction) of motion, rather in the sky directions $90^\circ$ away from the direction of local motion.

Several other cosmological probes have also been used to infer the motion of the Solar System with respect to \textit{the cosmological rest-frame}. One of the prominent probes is the number count of radio galaxies \cite{Blake:2002gx, Rubart:2013tx, Secrest:2020has}. Radio galaxies can be observed out to cosmological distances and hence the contamination due to sample variance in the local universe can be mitigated. Various works using data of radio sources available so far have provided differing estimates of the Solar System motion \cite{Rubart:2013tx}. However, current and next-generation radio surveys are poised to make an accurate and high significance detection \cite{Crawford:2008nh, Pant:2018smd, Bengaly:2018ykb, Nadolny:2021hti}. More recently, the effect of the Solar System motion on the thermal Sunyaev-Zeldovich effect \cite{zeldovich,sz1970} of galaxy clusters \cite{Chluba:2004vz} in \textit{Planck} data has been used to estimate the Doppler boost signal \cite{Akrami:2020nrk}. 

Correlations in the CMB introduced by the Doppler boost provide an equally competitive probe of the Solar System motion. Also, it is an important cross-check of the estimate of the Solar System motion from CMB Dipole and that of consistency of the CMB data-set. For high-resolution CMB maps, the detectability of such anisotropic imprints in the covariance matrix has been discussed in \cite{PhysRevLett.106.191301, Amendola:2010ty}. The earliest detection of such SI violation due to local motion, was done by the \textit{Planck} team \cite{Aghanim:2013suk} in 2013, using quadratic estimators proposed in \cite{Hanson_2009}. Consequently, various other estimators have been used to infer the Doppler boost signal \cite{Adhikari:2014mua, Aluri:2015tja}. The effect of these correlations on other quantities of interest in CMB studies is discussed in \cite{Catena:2012hq, Jeong_2014, Yasini:2019ajn}. A better understanding of the Doppler boost signal is also important for determining the possible contribution of the intrinsic CMB dipole \cite{Roldan:2016ayx, Meerburg:2017xga, Ferreira:2020aqa}. There are also work in the literature alluding to possible hints of an axis of evil \cite{Naselsky:2011jp, Zhao:2013jya}, which indicates strong directional alignment of the lowest multipoles CMB dipole, quadrupole and octopole along with the direction of parity asymmetry of CMB corresponding to  the largest angular scale. However, our inference of the local motion has been carried out from the effects of the motion on the CMB anisotropy at significantly smaller angular scales ($l_{\text{range}}=800-1950$). Hence, we can safely say that our analysis circumvents any possible contamination from the said effect and do not have implications for the same. 

In this work, we analyze \textit{Planck}-2018 \texttt{SMICA} temperature map \cite{Akrami:2018vks} in the BipoSH framework \cite{Mukherjee:2013zbi}. We do a joint Bayesian inference of the CMB power spectrum ($C_{l}$), the dipolar ($L=1$) BipoSH coefficients (which contain information about the off-diagonal terms caused by the local motion) and the temperature field in the spherical harmonic space ($a_{lm}$). As we discuss in Section \ref{DB_intro_and_cov}, such joint inference is obligatory owing to the fact that the non-SI off-diagonal part of the harmonic space covariance matrix is influenced by the SI diagonal part. Due to high dimensional parameter space, we employ the \textit{Hamiltonian Monte-Carlo} method \cite{Duane:1987de} to explore the posterior distribution of our model parameters i.e. the Doppler boost signal, $C_l$'s and $a_{lm}$'s of the CMB map. We use the inference formalism developed in \cite{SantanuDas, Shaikh:2019dvb}. We test our methodology on  simulated Doppler boosted CMB maps (generated using \texttt{CoNIGS} \cite{Mukherjee:2013kga}) and apply it to the \textit{Planck}-2018 \texttt{SMICA} temperature data. Being a frequency dependent effect, the Doppler boost signal affects data in different frequency channels differently. The \texttt{SMICA} CMB map is a combination of maps in the nine frequency channels. To deal with this complication, we introduce a new quantity, \texttt{SMICA} boost factor $b^{\texttt{SMICA}}(l)$ to extract the value of our local motion from the \texttt{SMICA} map.

We have organized the paper as follows. In section \ref{DB_intro_and_cov}, we review the Doppler boost model and provide a quantitative discussion of the modulation and aberration effects. We present the non-SI corrections of $a_{lm}$'s of the \texttt{SMICA} map for the Doppler boosted sky and their covariance matrix. In Section \ref{Method}, we set out the methodology of our approach. We discuss the posterior distribution of the parameters of interest and the details of the sampling method. In Section \ref{results}, we discuss some implementation details of our method and present the findings of our analysis on \textit{Planck}-2018 \texttt{SMICA} temperature map. Here, we also provide model comparisons for some known values of the signal strength in the literature, with the aid of the Bayes factor. Finally in Section \ref{Conclusion}, we discuss the summary of the work and the future outlook.

\section{Overview of the Doppler boost signal in the BipoSH framework}\label{DB_intro_and_cov}
In this section, we discuss how the statistical isotropy of the observed CMB fluctuations is violated in a boosted observation frame. In particular, we provide the expression for the harmonic space CMB covariance matrix in presence of the Doppler boost signal. The harmonic space CMB covariance matrix can be neatly expressed using the BipoSH representation, which we adopt throughout the paper. 

\subsection{Imprint of Doppler boost on CMB}
We denote the CMB temperature in the direction $\hat{n}'$ in the CMB rest frame by $T'(\hat{n'})$. Our observation frame, fixed with respect to the Solar System barycentre, is moving with the velocity $\bs{\beta}$ relative to the CMB rest frame.  The observed CMB temperature in the direction $\hat{n}$, $T(\hat{n})$ is given by the following expression \cite{Challinor:2002zh},
\begin{align}
    T(\hat{n}) = \frac{T'(\hat{n}')}{\gamma(1-\hat{n}\cdot\boldsymbol{\beta})},
\end{align}
where $\gamma = (1-\beta^2)^{-\frac{1}{2}}$ is Lorentz factor and $\beta \equiv |\bm{\beta}|$. The observed direction $\hat{n}$, in terms of $\hat{n}'$ and $\bm{\beta}$, is \cite{Challinor:2002zh, Amendola:2010ty, Aghanim:2013suk}
\begin{align}
    \hat{n} = \frac{\hat{n}'\cdot\hat{\beta} + \beta}{1+\hat{n}'\cdot\bs{\beta}}\hat{\beta} + \frac{\hat{n}' - (\hat{n}'\cdot\hat{\beta})\hat{\beta}}{\gamma(1+\hat{n}'\cdot\bs{\beta})}.
\end{align}
Expanding the observed temperature $T(\hat{n})$ up to linear order in $\beta$, the observed temperature fluctuations (excluding the CMB dipole) can be derived to be \cite{Aghanim:2013suk, Mukherjee:2013zbi}
\begin{align}
    \delta T(\hat{n}) = \delta T'(\hat{n}-\nabla(\hat{n}\cdot\boldsymbol{\beta}))(1+\hat{n}\cdot\boldsymbol{\beta}).
\end{align}
The \textit{Planck} detectors measure the intensity fluctuations at nine frequency channels. The intensity of CMB radiation at frequency $\nu$ is \cite{Kamionkowski:2002nd, Aghanim:2013suk}
\begin{align}
    I_{\nu}(\hat{n}) = \frac{2h\nu^3}{c^2}\frac{1}{\exp[h\nu /k_{B}T(\hat{n})]-1}.
\end{align}
Hence, the temperature fluctuation $\delta T(\hat{n})_{I}$ inferred from the intensity fluctuation $\delta I_{\nu}(\hat{n})$ at frequency $\nu$ is \cite{Aghanim:2013suk, Mukherjee:2013zbi}
\begin{align}\label{temp_flac_I}
    \delta T(\hat{n})_{I} = \frac{\delta I_{\nu}(\hat{n})}{dI_{\nu}/dT|_{T_{0}}} =  \delta T'(\hat{n}-\nabla(\hat{n}\cdot\boldsymbol{\beta}))(1 + b_{\nu}\hat{n}\cdot\boldsymbol{\beta}),
\end{align}
where the frequency dependent boost factor $b_{\nu}$ has the following form
\begin{align}
    b_{\nu} = \frac{\nu}{\nu_{0}}\coth\left(\frac{\nu}{2\nu_{0}}\right) - 1,
\end{align}
with $\nu_0 = 57$ GHz.

We carry out the whole analysis in the spherical harmonic space. For given $\bm{\beta}$, $D(\hat{n})$ represents the dipole field$^{\footnotemark}$\footnotetext{Since D($\hat{n}$) is a real-valued function, the coefficients $\beta_{11}$ and $\beta_{1-1}$ are related as $\beta_{1-1}^{\ast} = -\beta_{11}$.}, which has a form $\boldsymbol{\beta}\cdot\hat{n}$. In the spherical harmonic basis
\begin{align}\label{dipole_field}
    D(\hat{n}) \equiv \boldsymbol{\beta}\cdot \hat{n} = \sum_{N=-1}^{1} \beta_{1N}Y_{L=1,N}(\hat{n}),
\end{align}
where the peculiar velocity vector $\boldsymbol{\beta}$ has an amplitude $\beta$ and a direction $\hat{\beta} \equiv (\theta_{\beta}, \phi_{\beta})$ in the real space. For convenience, we use $\beta_{10}$, the real part of $\beta_{11}$ ($\beta_{11}^r$) and the imaginary part $\beta_{11}$ ($\beta_{11}^i$) as three real valued independent variables in our analysis. The real space variables, $\beta$, $\theta_{\beta}$ and $\phi_{\beta}$ are related to the spherical harmonic coefficients of the dipole field as \cite{Shaikh:2019dvb},
\begin{align}\label{real_var_relation}
    &\beta = \sqrt{\frac{3}{4\pi}}\sqrt{\beta_{10}^2 + 2\beta_{11}^{r^2} + 2\beta_{11}^{i^2}}, \nonumber\\
    &\theta_{\beta} = \cos^{-1}\left(\frac{\beta_{10}}{\beta}\sqrt{\frac{3}{4\pi}}\right), \nonumber\\
    &\phi_{\beta} = - \tan^{-1}\left(\frac{\beta_{11}^i}{\beta_{11}^r}\right).
\end{align}

 We assume that the CMB sky is statistically isotropic in its rest frame. We simplify \Eq{temp_flac_I} and calculate the $a_{lm}$'s for our boosted observation frame, in terms of the $\tilde{a}_{lm}$'s for the SI CMB rest frame. Up to  linear order in $\beta$, \cite{Mukherjee:2013zbi}
\begin{align}
    &\delta T(\hat{n}) = (1 + b_{\nu}\hat{n}\cdot\boldsymbol{\beta})(\delta T'(\hat{n}) - \nabla_{i}\delta T'(\hat{n})\nabla^{i}(\hat{n}\cdot\boldsymbol{\beta}) ).
\end{align}
Expressing the above equation in spherical harmonic space, we get
\begin{align}\label{alm_freq}
    a_{lm} = \tilde{a}_{lm} &+ b_{\nu}\sum_{N=-1}^{1}\beta_{1N}\sum_{l'm'}\tilde{a}_{l'm'}(-1)^{m}\frac{\Pi_{l'l}}{\sqrt{12\pi}}C_{l'0l0}^{10}C_{l'm'l-m}^{1N} \nonumber \\
    &-\sum_{N=-1}^{1}\beta_{1N}\sum_{l'm'}\tilde{a}_{l'm'}\frac{1}{2}[l'(l'+1)-l(l+1)+2](-1)^{m}\frac{\Pi_{l'l}}{\sqrt{12\pi}}C_{l'0l0}^{10}C_{l'm'l-m}^{1N},
\end{align}
where the spherical harmonic coefficients of the SI CMB sky are $\tilde{a}_{lm}=\int\delta T^{\text{SI}}(\hat{n})Y_{lm}(\hat{n})d\Omega_{\hat{n}}$. $C_{l'm'lm}^{LN}$ denote the Clebsch-Gordan coefficients and $\Pi_{l_1l_2...l_n} \equiv \sqrt{(2l_1+1)(2l_2+1)...(2l_n+1)}$ \cite{doi:10.1142/0270}. This is the expression of $a_{lm}$'s for a Doppler boosted CMB map at frequency $\nu$. Key aspects of the expression are that the non-SI corrections of $a_{lm}$'s due to local motion have a dependence on isotropic temperature field $\tilde{a}_{lm}$'s and the corrections are frequency dependent through the boost factor $b_{\nu}$. In the next section, we will exhibit how these non-SI corrections generate non-zero off-diagonal terms in the harmonic space CMB covariance matrix.

\subsection{Covariance matrix of Doppler boosted CMB temperature map}\label{DB_model}
\begin{figure}
\centering
\includegraphics[width=38pc]{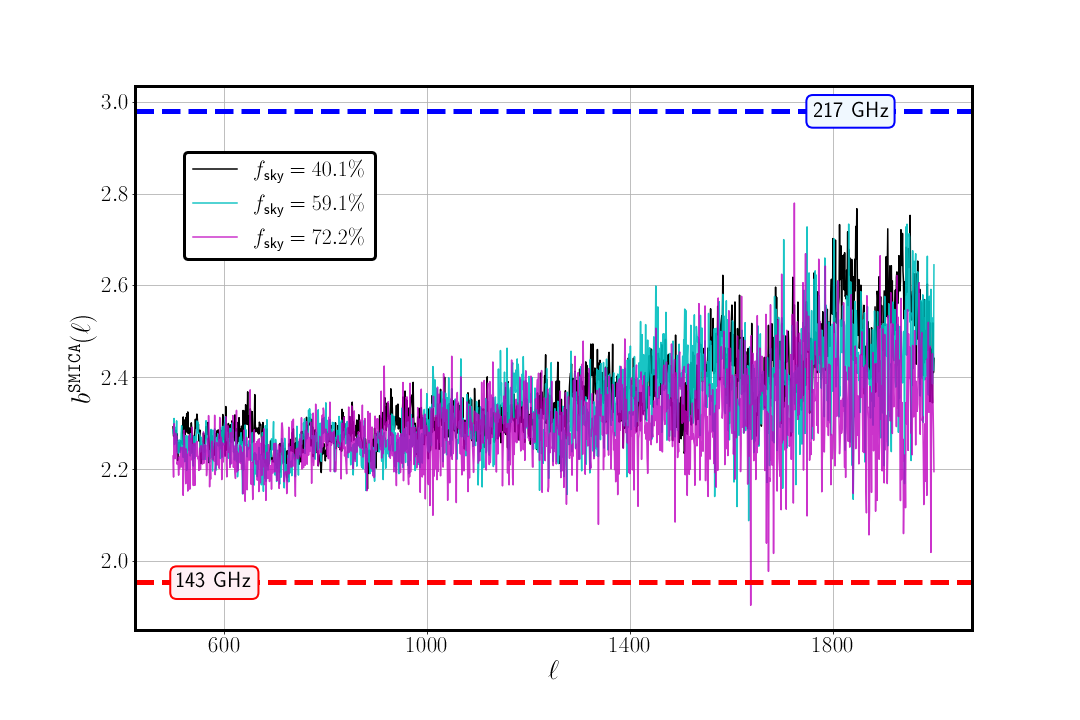}
\caption{ In this figure we show the \texttt{SMICA} boost factor $b^{\texttt{SMICA}}(l)$ for 3 different masks, $f_{\text{sky}}=40.1\%$ (black curve), $59.1\%$ (cyan curve) and $72.2\%$ (magenta curve), which are depicted in Figure \ref{fig_map1}. $b^{\texttt{SMICA}}(l)$ corresponding to each mask is obtained by taking the peaks of histograms of $b_{lm}^{\texttt{SMICA}}$ (we have $(2l+1)$ modes at each $l$). As a reference, the red and the blue lines show the boost factor $b_{\nu}$ for the \textit{CMB channels}, 143 GHz and 217 GHz, respectively.}\label{eff_bnu}
\end{figure}

We use the \texttt{SMICA} estimate of the CMB temperature anisotropy map provided by \textit{Planck}. The \texttt{SMICA} method makes use of maps in all  nine frequency channels \cite{Delabrouille:2002kz, 4703509}. However, for the \textit{Planck}-2018 SMICA map, not all the data from all the channels are used for the whole sky. Though \texttt{SMICA} is a harmonic space method, for the \textit{Planck} 2018 release, it makes use of a real space filter in the form of a mask along with a harmonic space filter \cite{Akrami:2018mcd}. The \texttt{SMICA} map provided in the 2018 data release of the \textit{Planck} is constructed as summarised in the following expression \cite{Akrami:2018mcd}
\begin{align}\label{map_smica}
    X_{\texttt{SMICA}} = X_{\text{full}} + P(X_{\text{high}}-X_{\text{full}}),
\end{align}
where $X_{full}$ and $X_{high}$ have the following form,
\begin{align}
    X_{\text{full}}(\hat{n}) = \sum_{\nu}^{\text{full}}\sum_{lm} a_{lm}^{\nu}Y_{lm}(\hat{n})W_{\nu}^{\text{full}}(l), \nonumber \\
    X_{\text{high}}(\hat{n}) = \sum_{\nu}^{\text{high}}\sum_{lm} a_{lm}^{\nu}Y_{lm}(\hat{n})W_{\nu}^{\text{high}}(l).
\end{align}
In the above expressions, ``full'' stands for all the nine \textit{Planck} frequency channels and ``high" stands for only the six HFI frequency channels. The weights $W_{\nu}^{\text{full}}(l)$ and $W_{\nu}^{\text{high}}(l)$ are defined and given by \textit{Planck}\footnote{\texttt{weights\_T\_smica\_R3.00\_Xfull.txt} and \texttt{weights\_T\_smica\_R3.00\_Xhigh.txt} in \texttt{COM\_Code\_SMICA-weights-propagation\_R3.00.tar.gz} available in \url{https://wiki.cosmos.esa.int/planck-legacy-archive/index.php/SMICA_propagation_code}}. The operator ${P}$ in \Eq{map_smica}, is a hybridization of two operators. One is an apodized galactic mask ${M}$\footnote{\texttt{transition\_mask.fits.gz} in \texttt{COM\_Code\_SMICA-weights-propagation\_R3.00.tar.gz}}, the other one is an apodized high-pass filter $\mathcal{F}$ in harmonic space. $\mathcal{F}$ has the following functional form\footnote{\label{note1}See \texttt{smica\_coadd.py} in \texttt{COM\_Code\_SMICA-weights-propagation\_R3.00.tar.gz}},
\begin{align}
    \mathcal{F} &= 0 &\text{for $l < l_1$}, \nonumber \\
      &=\frac{1}{2} - \frac{1}{2}\cos{\left[\frac{\pi(l-l_1)}{l_2 - l_1}\right]} \quad &\text{for $l_1 \leq l \leq l_2$}, \nonumber \\
      &= 1  &\text{for $l > l_2$}.
\end{align}
So the combined effect of the operator ${P}$ is as follows. In masked regions of CMB sky, where ${M}= 0$, $X_{\texttt{SMICA}} = X_{\text{full}}$. In unmasked regions, where ${M}=1$, for $l < l_1$, $X_{\texttt{SMICA}} = X_{\text{full}}$ and for $l>l_2$, $X_{\texttt{SMICA}} = X_{\text{high}}$. In the \texttt{SMICA} pipeline, $l_1= 50$ and $l_2 = 150$ values are used\textsuperscript{\ref{note1}}. As long as we consider regions where $M=1$ and $l_{\text{range}}$ over $150$, it is safe to approximate $X_{\texttt{SMICA}} = X_{\text{high}}$. The $a_{lm}$'s of \texttt{SMICA} map ($ a_{lm}^{\texttt{SMICA}}$), in terms of the $a_{lm}$'s of the single frequency maps ($a_{lm}^{\nu}$) is then
\begin{align}\label{alm_smica}
    a_{lm}^{\texttt{SMICA}} = \sum_{\nu}^{\text{high}}W_{\nu}^{\text{high}}(l)a_{lm}^{\nu}.
\end{align}
Using the expression in \Eq{alm_freq} of the $a_{lm}$'s for the boosted single channels in \Eq{alm_smica}, we get the expression of $a_{lm}$'s of boosted \texttt{SMICA} map 
\begin{align}\label{alm_smica_DB}
    a_{lm}^{\texttt{SMICA}} = \tilde{a}_{lm}^{\texttt{SMICA}} &+ b^{\texttt{SMICA}}_{lm}\sum_{N=-1}^{1}\beta_{1N}\sum_{l'm'}\tilde{a}_{l'm'}^{\texttt{SMICA}}(-1)^{m}\frac{\Pi_{l'l}}{\sqrt{12\pi}}C_{l'0l0}^{10}C_{l'm'l-m}^{1N} \nonumber \\
    &-\sum_{N=-1}^{1}\beta_{1N}\sum_{l'm'}\tilde{a}_{l'm'}^{\texttt{SMICA}}(-1)^{m}\frac{1}{2}[l'(l'+1)-l(l+1)+2]\frac{\Pi_{l'l}}{\sqrt{12\pi}}C_{l'0l0}^{10}C_{l'm'l-m}^{1N},
\end{align}
which is a reduced form similar to a single channel expression, but with a boost factor $b_{lm}^{\texttt{SMICA}}$, given by
\begin{align}\label{b_eff_nu_lm}
     b_{lm}^{\texttt{SMICA}} \equiv \frac{\sum_{\nu}^{\text{high}}b_{\nu}W_{\nu}^{\text{high}}(l)a_{lm}^{\nu}}{a_{lm}^{\texttt{SMICA}}}.
\end{align}
We calculate this $b_{lm}^{\texttt{SMICA}}$ from the six HFI frequency channels, with the aid of publicly available \texttt{SMICA} propagation code\textsuperscript{\ref{note1}}. We choose the most probable values from the $b_{lm}^{\texttt{SMICA}}$ distributions of $(2l+1)$ modes at every multipole $l$, to obtain $b^{\texttt{SMICA}}(l)$\footnote{We avoid using the mean value of $b_{lm}^\texttt{SMICA}$, as it is going to be driven by the extended tail of the  distribution.}. The value chosen as  $b^{\texttt{SMICA}}(l)$ makes sure that it is a representative values for most of the $b_{lm}^{\texttt{SMICA}}$ values at a fixed $l$. The non-SI corrections of $a_{lm}^{\texttt{SMICA}}$ due to local motion in \Eq{alm_smica_DB} are constituted from all the quantities that are accessible to us through the $\texttt{SMICA}$ map. For our analysis, we compute $b^{\texttt{SMICA}}(l)$ for three different choices of masks shown in Figure \ref{eff_bnu}. The increase in the value of  $b^{\texttt{SMICA}}(l)$ at high $l$ arises due to larger value of the \texttt{SMICA} weights for the frequency channel $217$ GHz at values of $l\geq 1600$ \cite{Akrami:2018mcd}.

We use \Eq{alm_smica_DB} to compute the covariance matrix elements for the \texttt{SMICA} map due to the Doppler boost. Up to leading order in $\beta$,\footnote{$\langle .\rangle$ denotes the ensemble average.}
\begin{align}\label{cov_mat}
    \langle a_{lm}a_{l'm'}^{\ast} \rangle = &C_{l}\delta_{ll'}\delta_{mm'} \nonumber \\
    + &(-1)^{m'}\frac{\Pi_{ll'}}{\sqrt{12\pi}}\sum_{N=-1}^{1}\beta_{1N}[(b^{\texttt{SMICA}}(l)-\frac{1}{2}\{l(l+1)-l'(l'+1)+2\})C_{l} \nonumber \\
    + &(b^{\texttt{SMICA}}(l')-\frac{1}{2}\{l'(l'+1)-l(l+1)+2\})C_{l'}]C_{l0l'0}^{10}C_{lml'-m'}^{1N}.
\end{align}
A few important aspects of this expression are as follows. The first term in \Eq{cov_mat}, provides the diagonal elements $C_l$'s, which come from the $\tilde{a}_{lm}^{\texttt{SMICA}}$'s in \Eq{alm_smica_DB}. The second term in \Eq{cov_mat}, represents the leading order off-diagonal terms, which are linear in $\beta$. If the velocity of our local motion $\beta$ were zero, then the $\beta_{1N}$'s would have been zero and we would have been left with a diagonal covariance matrix as expected for a SI sky. The off-diagonal part has a dependence on the $C_l$'s, which again reinforces the point that joint analysis of local motion variables along with the $C_l$'s, is essential. There is a correction in the diagonal terms as well, if we include the next order for $\beta$ in our calculation. We have ignored the corrections to $C_l$ due to local motion in this analysis, as it can lead to a maximum relative difference in its value by about a few times $10^{-3}$-- $10^{-4}$ for the partial-sky analysis relevant for this work (with a minimum $f_{\text{sky}}=40.1\%$) \cite{2011MNRAS.415.3227C,2013MNRAS.432.2208Y,Jeong_2014, Yasini:2019ajn}. These values are sub-dominant in comparison to the total uncertainties (cosmic variance + Planck instrument noise \cite{Ade:2015via}) in the value of $C_l$ at high multipole values considered in this analysis \cite{Akrami:2018vks}. However, for future high-resolution and more sensitive CMB experiments, higher-order terms will play a crucial role and cannot be ignored \cite{Jeong_2014, Yasini:2019ajn}.
 
Any particular cause behind the violation of SI of the CMB manifests through non-zero values of only certain off-diagonal terms in the covariance matrix. BipoSH coefficients ($A^{LN}_{l_{1}l_{2}}$) provide an elegant way to group these particular terms according to the nature of the isotropy violating correlations, organizing them according to how they transform under rotations \cite{VMK}. The elements of the covariance matrix can be written as a linear combination of the BipoSH coefficients,
\begin{align}\label{biposh_1}
    \langle a_{lm}a_{l'm'}^{\ast} \rangle = (-1)^{m'}\sum_{LN}A^{LN}_{ll'}C_{lml'-m'}^{LN}.
\end{align}
The BipoSH coefficients form a complete basis for expressing all the terms in the covariance matrix. In particular, the diagonal elements of the covariance matrix are just the $L=0,\, M=0$ terms of the BipoSH coefficients, given by
\begin{align}
    A_{l l'}^{00} = (-1)^{l}\sqrt{2l + 1}C_{l}\delta_{l l'}.
\end{align}
The Clebsch-Gordon coefficients in \Eq{cov_mat} limit the number of off-diagonal terms we get in the covariance matrix for a boosted sky. We get only the mixing of $l$ and $l\pm1$ terms. Therefore the diagonal terms and just the immediately adjacent terms on both sides of the diagonal terms in the covariance matrix are non-zero. As the BipoSH coefficients for this case are symmetric under the exchange of lower indices, $l$ and $l+1$, we only need the $A^{1N}_{l,l+1}$ terms to express all the off-diagonal terms. We compare \Eq{cov_mat} and \Eq{biposh_1} to get the expression of the BipoSH coefficients as
\begin{align}\label{ALM}
    A^{1N}_{l,l+1} = \beta_{1N}\mathcal{S}_{l,l+1},
\end{align}
where $\beta_{1N}$ is the amplitude of the Doppler boost signal and $\mathcal{S}_{l,l+1}$ is called the \textit{shape factor} which is defined as,
\begin{align}\label{sh_fac}
    \mathcal{S}_{l,l+1} \equiv \frac{\Pi_{l,l+1}}{\sqrt{12\pi}}[(l+b^{\texttt{SMICA}}(l))C_l - (l+2-b^\texttt{SMICA}(l+1))C_{l+1}]C^{10}_{l0l+10}.
\end{align}
We use this particular form of the shape factor in our analysis to extract the Doppler boost signal from the \textit{Planck}-2018 \texttt{SMICA} temperature map \cite{Akrami:2018vks}.

\section{Method}\label{Method}

\subsection{Data Model and the Parameter Posterior}\label{post_dist}
We use the following model for the data $\bs{d}$ \footnote[8]{We denote vectors and matrices by bold symbol.},
\begin{equation}\label{eq_data_model}
 \bs{d} = \boldsymbol{\mathcal{B}}\bs{s} + \bs{N},
\end{equation}
where $\bs{s}$ is the CMB signal vector, $\bs{N}$ is the noise vector and $\boldsymbol{\mathcal{B}}$ is the beam convolution operator. We assume the noise to be Gaussian \cite{Akrami:2018vks, Aghanim:2018fcm}. Hence, the probability of the data $\bs{d}$ given signal $\bs{s}$ is
\begin{equation}\label{eq_prob_d_s}
 \calP(\textbf{d}|\bs{s}) = \frac{1}{\sqrt{|2 \pi \bs{C_N}|} } \exp \Big[ -\frac{1}{2} 
(\bs{d} - \bs{s})^{\dagger} \bs{C_N}^{-1} (\bs{d} - \bs{s}) \Big ],
\end{equation}
where, $\bs{C_N}$ is the noise covariance matrix. We want to sample the joint posterior distribution of the parameters $a_{lm}$, $C_l$ and $\beta_{1N}$. Using the probability distributions given above, we can express the posterior distribution of $a_{lm}$, $C_l$ and $\beta_{1N}$ as
\begin{equation}\label{eq_Bayes_expr1}
 \calP(\bs{C_S}, \bs{s} | \bs{d}) = \frac{ \mathcal{L}(\bs{d} | \bs{s}, \bs{C_S}) 
 \Pi(\bs{s}, \bs{C_S})}{\mathcal{E}(\bs{d})},
\end{equation}
where $\mathcal{L}$ is the likelihood, $\mathcal{E}$ is the evidence and $\Pi$ is the prior. For the likelihood, conditioned on $\bs{s}$, $\bs{d}$ is independent of $\bs{C_S}$, implying $\mathcal{L}(\bs{d} | \bs{s}, \bs{C_S}) = \mathcal{L}(\bs{d} | \bs{s})$. 
Hence, we have
\begin{equation}\label{eq_prob_CS_s}
 \calP(\bs{C_S}, \bs{s} | \bs{d}) = \frac{ \mathcal{L}(\bs{d} | \bs{s}) 
\Pi(\bs{s}|\bs{C_S}) \Pi(\bs{C_S})}{\mathcal{E}(\bs{d})},
\end{equation}
where we used $\Pi(\bs{s}, \bs{C_S}) =  \Pi(\bs{s}|\bs{C_S}) \Pi(\bs{C_S})$.
For the prior on the signal amplitude $\Pi(\bs{s}|\bs{C_S})$, we use the theoretically motivated and empirically established fact that the Gaussianity of the CMB fluctuations is a very good approximation \cite{Ade:2013nlj,Ade:2015hxq, Akrami:2019bkn, Akrami:2019izv}. Hence, 
\begin{equation}\label{eq_prob_s_CS}
\Pi(\bs{s}|\bs{C_S}) = \frac{1}{\sqrt{|2 \pi \bs{C_S}|}} \exp \Big[ -\frac{1}{2} 
\bs{s}^{\dagger} \bs{C_S}^{-1} \bs{s} \Big ].
\end{equation}
In the above equations, $\mathcal{E}(\bs{d})$ is the Evidence of the data and is a normalization constant. The explicit expression for the Evidence for this problem would be
\begin{equation}
\mathcal{E}(\bs{d}) = \int \bs{dC_S} \bs{ds} \mathcal{L}(\bs{d} | \bs{s}) 
\Pi(\bs{s}|\bs{C_S}) \Pi(\bs{C_S}).
\end{equation}
Evidence is generally computed for the purpose of model comparison. However, for the  probability distributions we consider, brute force computation of the evidence is intractable. Instead, we bypass explicit computation of the above integration by using the Savage Dickey Density Ratio \cite{10.1214/aoms/1177693507, Trotta:2008qt} to perform model comparison in Section \ref{results}. $\Pi(\bs{C_S})$ is now the prior on the CMB signal covariance matrix and we assume it to be flat. The robustness of analysis for different choices of prior has been discussed in \cite{Shaikh:2019dvb}, for similar scenario. Hence, the probability distribution to be sampled has the same functional form as $\calP(\bs{d} | \bs{s}) 
\calP(\bs{s}|\bs{C_S})$, explicitly expressed as
\begin{equation}\label{eq_post_s_CS}
 \calP(\bs{C_S}, \bs{s} | \bs{d}) = \frac{1}{\sqrt{|2 \pi \bs{C_N}| |2 \pi \bs{C_S}| } } \exp { \Big\{ -\frac{1}{2} 
 \Big[ (\bs{d} - \bs{s})^{\dagger} \bs{C_N}^{-1}  (\bs{d} - \bs{s}) + \bs{s}^{\dagger} \bs{C_S}^{-1} \bs{s} \Big ] \Big\} }.
\end{equation}
Note that the dependence of the posterior distribution on the parameters $a_{lm}$, $C_l$ and $\beta_{1N}$ is through the signal vector $\bs{s}$ and the signal covariance matrix $\bs{C_S}$.
The above formalism presents the Bayesian hierarchical model for the  problem at hand. The $a_{lm}$ form one set of model parameter which are directly compared with the data. Parameters of the signal covariance matrix form another set of parameters at a different level of hierarchy.

Given the large number of parameters, we use the Hamiltonian Monte Carlo (HMC) method \cite{Duane:1987de, 2011hmcm.book..113N} to jointly sample the posterior distribution $\mathcal{P}(\bs{C_S}, \bs{s} | \bs{d})$. The details of the sampling of this particular distribution, using the language of BipoSH coefficients to describe the covariance matrix, are discussed in \cite{SantanuDas, Shaikh:2019dvb}. Details of the particular methodology used here are discussed in \cite{Shaikh:2019dvb} in the context of dipole modulation (cosmic hemispherical asymmetry) of CMB temperature anisotropy \cite{Hoftuft_2009, Ade:2013nlj, Ade:2015hxq, Akrami:2019bkn}. We adopt this formalism for the inference of the Doppler boost signal. In the next section, we describe HMC in brief and provide some of the relevant mathematical expressions.

\subsection{Sampling method: Hamiltonian Monte Carlo}\label{HMC}
HMC is an efficient Monte Carlo sampling method and makes use of the Hamiltonian dynamics to propose the sample \cite{Duane:1987de, 2011hmcm.book..113N}. Further, HMC avoids the curse of dimensionality that affects  the Metropolis algorithm, especially for high dimensional inference and therefore potentially achieves much higher acceptance probability for proposed samples in the chain. HMC has been used in  cosmology research for various challenging inference problems with hierarchical Bayesian models; some examples are cosmological parameter estimation \cite{Hajian_HMC}, CMB power spectrum inference \cite{Taylor_Cl_with_HMC_2008}, various inference problems in large scale structure \cite{Jasche_2010, Jasche_2013, Jasche_Wandelt2013}, and inference of CMB lensing potential \cite{Anderes_Wandelt_Lavaux_bayes_lensing_2015, Millea:2020cpw}.

The Hamiltonian is defined as
\begin{equation}\label{eq_Hamiltonian}
\calH(\{q_i, p_i\}) = \frac{1}{2} \bs{p}^T \bs{\mu}^{-1} \bs{p} - \ln[\mathcal{P}(\bs{q})], 
\end{equation}
where $\bs{q}$ is a vector of the parameters of interest, $\bs{p}$ is a vector of the momentum associated with the parameters and $\bs{\mu}$ is the \textit{mass matrix}. The term $-\ln[\mathcal{P}(\bs{q})]$ is called the \textit{potential energy} for the distribution being sampled, here denoted by $\mathcal{P}(\bs{q})$. For our problem, $\mathcal{P}(\bs{q})$ is given by the expression in \Eq{eq_post_s_CS}. Hamilton's equations are
\begin{equation}
 \dot{q} \equiv \frac{dq}{dt} = \frac{\pd \calH}{\pd p} \quad \text{and} \quad
 \dot{p} \equiv \frac{dp}{dt} = -\frac{\pd \calH}{\pd q}.
\end{equation}
The main calculation involved in HMC concerns the evaluation of the momentum derivative for the parameters of interest. Hence, one needs to get the following quantity for all the parameters of interest
\begin{equation}
 \dot{p_j} \equiv \frac{dp_j}{dt} = -\frac{\pd \calH}{\pd q_j} = 
 \frac{\pd \ln[\mathcal{P}(\{q_i\})] }{\pd q_j}. 
\end{equation}
For our problem, the expression for the momentum derivative can be obtained analytically. In the remaining section, we present calculations of $\frac{\pd \ln[\mathcal{P}(\{q_i\})]}{\pd q}$, with $q = \{a_{lm}, C_l, \beta_{1N}\}$. 

The momentum derivative corresponding to $a_{lm}$ is given by
\begin{equation}\label{palm_dot_2}
  \dot{p}_{lm} = -\frac{1}{2} \sum_{l_1 m_1} [\bs{C_S}^{-1}]_{l_1m_1lm} a^{*}_{l_1m_1} + 
 \frac{1}{2} \sum_{l_1 m_1}[\bs{C_N}^{-1}]_{l_1m_1lm} (d^{*}_{l_1m_1} - a^{*}_{l_1m_1}).
\end{equation}
To deal with the mask, we express the probability distribution of the data given signal using real space representation of the data and the signal. The joint likelihood for the signal in $N_{pix}$ pixels is
\begin{equation}
 \calP(\bs{s} | \bs{d}) = \frac{1}{|2 \pi \bs{C_N}|^{1/2}} 
 \exp { \Big[ -\frac{1}{2} \sum^{N_{pix}}_{i = 1} \frac{(d_i - \mathcal{B}s_i)^{2}}{\sigma^{2}_{i} } \Big]},
\end{equation}
where $\mathcal{B}s_i$ is the signal in the $i^{th}$ pixel of the map smoothed with the beam $\mathcal{B}$ and $\sigma^2_i$ is the noise variance in the $i^{th}$ pixel. To accommodate the presence of mask, we express $\dot{p}_{lm}$ in the following form
\begin{equation}\label{plm_mask}
\dot{p}_{lm} = -\frac{1}{2} \sum_{l_1 m_1} S^{-1}_{l_1m_1lm} a^{*}_{l_1m_1} 
+ \frac{1}{2} \sum^{N_{pix}}_{i = 1} \frac{1}{\sigma^2_i} \Big[ \sum_{l'm'} 
(d_{l'm'} - b_l p_l a_{l'm'})b_l p_l Y_{l'm'}(\hat{n_i}) \Big]Y_{lm}(\hat{n_i}).
\end{equation}
The momentum derivative corresponding to the BipoSH coefficient $A^{LM}_{ll'}$ is \cite{Shaikh:2019dvb}
\begin{eqnarray}\label{PALMll_dot_2}
 \dot{p}^{LN}_{ll'} = &-& \frac{1}{2} \sum_{m'} \frac{(-1)^{m'}}{D_{l'm'l'm'}} C^{LN}_{lm'l'-m'} 
\delta_{ll'} \nonumber\\
 &-& \frac{1}{2} \sum_{m', m} \frac{(-1)^{m'} O_{l'm'lm} }{D_{l'm'l'm'} D_{lmlm}} C^{LN}_{lml'-m'}
 + \frac{1}{2} \sum_{m', m} \frac{(-1)^{m'} a_{l'm'} a^*_{lm} }{D_{l'm'l'm'} D_{lmlm}} 
C^{LN}_{lml'-m'}.
\end{eqnarray}
In particular, the momentum derivative for the BipoSH coefficient with $L = 0, M = 0$ is
\begin{equation}\label{eq_PA00ll_dot}
 \dot{p}^{00}_{ll'} = 
 \frac{2l + 1}{2A^{00}_{ll}} (\frac{\hat{A}^{00}_{ll'}}{A^{00}_{ll}} - 1),
 \quad \text{where} \quad 
 \hat{A}^{00}_{ll'}  = \sum_{m m'} a_{lm} a^{*}_{l'm'} C^{00}_{lml'm'}.
\end{equation}
The momentum derivative with respect to $\beta_{1N}$ parameter is
\begin{equation}\label{Pm1M_dot}
\frac{\pd \calH}{\pd \beta_{1N}} = 
\sum_{l} \Big[ \frac{\pd A^{1N}_{l l+1}}{\pd \beta_{1N}} \frac{\pd \calH}{\pd A^{1N}_{ll+1}} + \frac{\pd A^{1N}_{l+1 l}}{\pd \beta_{1N}} \frac{\pd \calH}{\pd A^{1N}_{l+1l}} \Big].
\end{equation}
The momentum derivatives for the real and imaginary parts of $\beta_{11}$ are obtained using the following expressions
\begin{equation}
 \frac{\pd \calH}{\pd \beta^r_{11}} = 2\Re \Big[ \frac{\pd \calH}{\pd \beta_{11}} \Big]
 \quad \text{and} \quad
 \frac{\pd \calH}{\pd \beta^i_{11}} = -2\Im \Big[ \frac{\pd \calH}{\pd \beta_{11}} \Big],
\end{equation}
where $\Re$ and $\Im$ are operators giving real and imaginary parts of the expression, respectively.

\section{Results: Estimation of the local motion from \textit{Planck}-2018 CMB temperature map}\label{results}
This section is devoted to the results of our analysis on \textit{Planck}-2018 \texttt{SMICA} temperature map\footnote[9]{\texttt{COM\_CMB\_IQU-smica\_2048\_R3.00\_full.fits} available at \url{https://pla.esac.esa.int/}}. In the first subsection, we lay out some implementation details of our analysis on \texttt{SMICA} 2018 temperature map, like masking, noise etc. In the subsequent sections, we present the result of our analysis and compare it with some known values in the literature.
\begin{figure}
\centering
\subfigure[ ]{\label{fig_map1_a}  
\includegraphics[width=17pc]{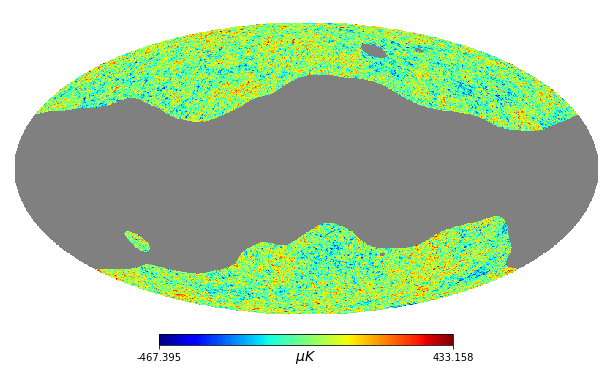}}\hspace{7mm}
\subfigure[ ]{\label{fig_map1_b} 
\includegraphics[width=17pc]{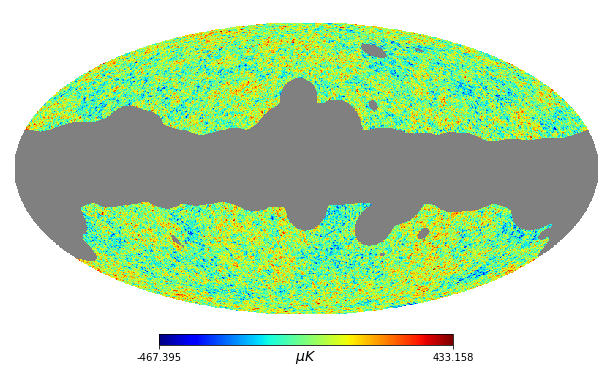}}
\subfigure[ ]{\label{fig_map1_c}  
\includegraphics[width=17pc]{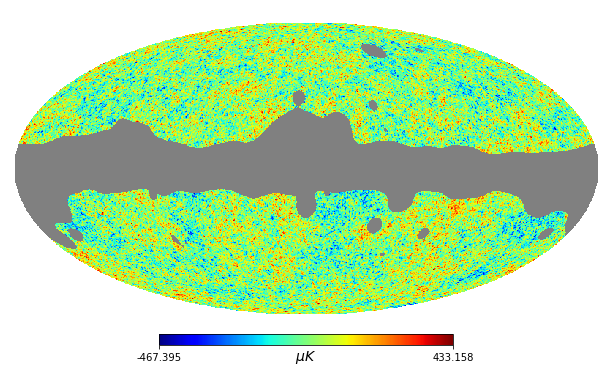}}\hspace{7mm}
\subfigure[ ]{\label{fig_map1_d} 
\includegraphics[width=17pc]{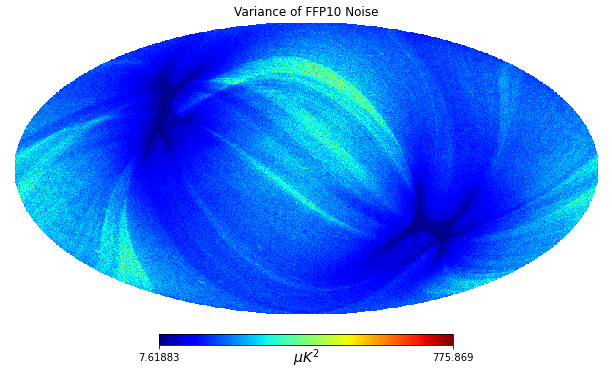}}
\caption{The grey regions in Figure \ref{fig_map1_a}, \ref{fig_map1_b} and \ref{fig_map1_c} are the three different masks ($f_{\text{sky}}=40.1\%$, $59.1\%$ and $72.2\%$ respectively), which have been used in our analysis. Figure \ref{fig_map1_d} is the noise variance map, obtained from 200 FFP10 noise simulations, that has been used in our analysis.}\label{fig_map1}
\end{figure}\label{fig_maps1}
\subsection{Implementation}
As discussed in Section \ref{DB_model}, the \texttt{SMICA} pipeline provides us a hybrid map from the nine \textit{Planck} frequency channels. The \texttt{SMICA} map is constructed in a specific manner \cite{Akrami:2018mcd} to minimize the contamination from foregrounds. Hence, as long as we are considering the cleaner regions of the CMB sky well away from the galactic plane and using multipoles higher than $l=150$ for our analysis, we can use the approximation given in \Eq{alm_smica}. Accordingly, we have performed our analysis on \textit{Planck}-2018 \texttt{SMICA} temperature map for three different choices of masking and three different choices of the maximum CMB  multipoles, which we have discussed elaborately later. These three masks have been given in Figure \ref{fig_map1_a}, \ref{fig_map1_b} and \ref{fig_map1_c}. We need the noise variance in each pixel to be used in \Eq{plm_mask}. To obtain an estimate of the noise variance, we use 200 FFP10 simulations of the \texttt{SMICA} noise \cite{Ade:2015via}. At every pixel, we compute the variance of 200 FFP10 simulated noise maps\footnote[10]{\url{https://pla.esac.esa.int/}} stored as \texttt{HEALPix} \cite{Gorski:2004by} maps with $N_{\text{side}}$ = 2048. The noise variance map is provided in Figure \ref{fig_map1_d}. A more accurate computation of noise covariance matrix in pixel-space can be performed using a recipe developed in \cite{Ramanah:2019rpg} and will be used in a future work. For the masked pixels excluding regions with high galactic foreground contamination and pixels at high-latitudes that coincide with known point source locations, we assign infinite variance by making $1/\sigma^2 = 0$ in \Eq{plm_mask}. This is the equivalent of having no information at those pixels. The anisotropy of the  noise is encoded in the noise variance map. For the calculation of the mass matrix of $a_{lm}$'s \cite{Taylor_Cl_with_HMC_2008}, the noise power spectrum $N_{l}$ is obtained from the difference of the two \texttt{SMICA} 2018 half-mission maps\footnote[11]{\texttt{COM\_CMB\_IQU-smica\_2048\_R3.00\_hm1.fits} and \texttt{COM\_CMB\_IQU-smica\_2048\_R3.00\_hm2.fits} available at \url{https://pla.esac.esa.int/}}, using the fact that noises of the two half-mission maps are uncorrelated. We have used a smooth $N_{l}$ (with $l_{\text{width}} = 40$) in our analysis. In this joint inference approach, we also smooth the auxiliary $C_l$ using a rectangular window function of bin width  $l_{\text{width}}=40$ and apply it in momentum derivative expressions of BipoSH, given in \Eq{PALMll_dot_2}.

The whole analysis has been carried out at a resolution $N_{\text{side}} = 2048$, which grants us sufficiently high $l_{\text{range}}$. The modulation part of the Doppler boost signal is degenerate with the dipolar modulation signal of cosmic hemispherical asymmetry \cite{Hoftuft_2009, Ade:2013nlj, Ade:2015hxq, Akrami:2019bkn, Shaikh:2019dvb} at low $l$. Hence we restrict the multipoles used in our analysis down to $l_{\text{min}}=800$ to circumvent possible contamination in the estimation of $\beta$. We carry out our analysis using two different $l_{\text{range}}$ from $l_{\text{min}}=800$, with $l_{\text{max}}=1500 \text{ and } 1950$\footnote[12]{We have also obtained the result for $l_{\text{max}}=1700$. It agrees well with the other $l_{\text{max}}$ choices.}. Since the Doppler boost signal is scale-independent and the variance of the $\beta_{1N}$ is inversely proportional to the number of modes considered \cite{Mukherjee:2013zbi, Shaikh:2019dvb}, it is crucial to perform the analysis up to high values of the multipoles $l$. The analysis using the mask $f_{\text{sky}}=40.1\%$, with $l_{\text{max}}=1950$ is our primary set-up in this work as it is the most conservative choice of a mask to reduce the contamination from galactic foregrounds and in getting a more reliable inference of the Doppler boost signal. In the next subsection, we have presented the detailed results corresponding to the primary setup. We also provide the inferred values from the other setups as well. 

\begin{figure}[t!]
\centering
\subfigure[ ]{\label{fig_map2_a}  
\includegraphics[width=17pc]{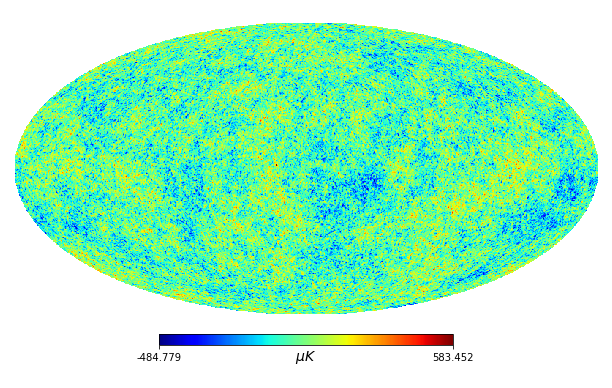}}\hspace{7mm}
\subfigure[ ]{\label{fig_map2_b} 
\includegraphics[width=17pc]{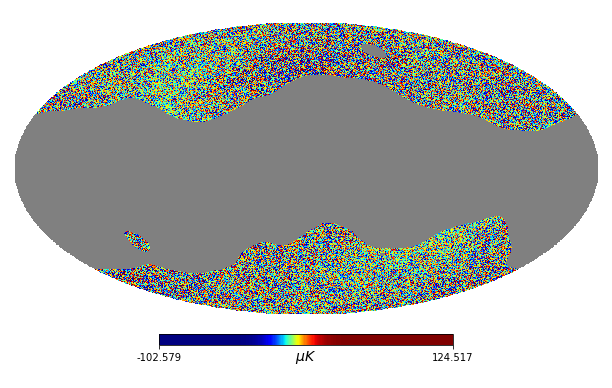}}
\caption{Figure \ref{fig_map2_a} shows one of the sample maps ($N_{\text{side}}=2048$) obtained from the $a_{lm}$ samples with the primary set-up ($f_{\text{sky}}=40.1\%$, $l_{\text{max}}=1950$). Figure \ref{fig_map2_b} shows the difference between the \texttt{SMICA} map and the sample map at unmasked regions.}
\end{figure}\label{fig_maps2}

\subsection{Result}
In this section, we present the findings of our analysis on \textit{Planck}-2018 \texttt{SMICA}  temperature map with the setup mentioned in the previous section. As discussed in Section \ref{post_dist}, our analysis provides a joint inference of the spherical harmonic coefficients ($a_{lm}$'s), the power spectra ($C_l$'s) and the Doppler boost signal ($\beta_{10}$, $\beta_{11}^r$ and $\beta_{11}^i$), as we explore the posterior distribution given in \Eq{eq_post_s_CS}, for \texttt{SMICA} map. 

We have obtained $5\times10^{4}$ samples of the model parameters using HMC sampling method discussed in Section \ref{HMC}. We discard the first 4000 samples as burn-in. In Figure \ref{fig_map2_a}, we show one of the samples of $a_{lm}$. In Figure \ref{fig_map2_b}, we show the difference between the sample map and the input \texttt{SMICA} map over the unmasked regions. From this figure, it is very prominent that the sample map and the input map agrees well, in the regions with less noise variance (see Figure \ref{fig_map1_d}).

\begin{figure}
\centering
\subfigure[ ]{\label{cl_plot_a}  
\includegraphics[width=0.9\linewidth]{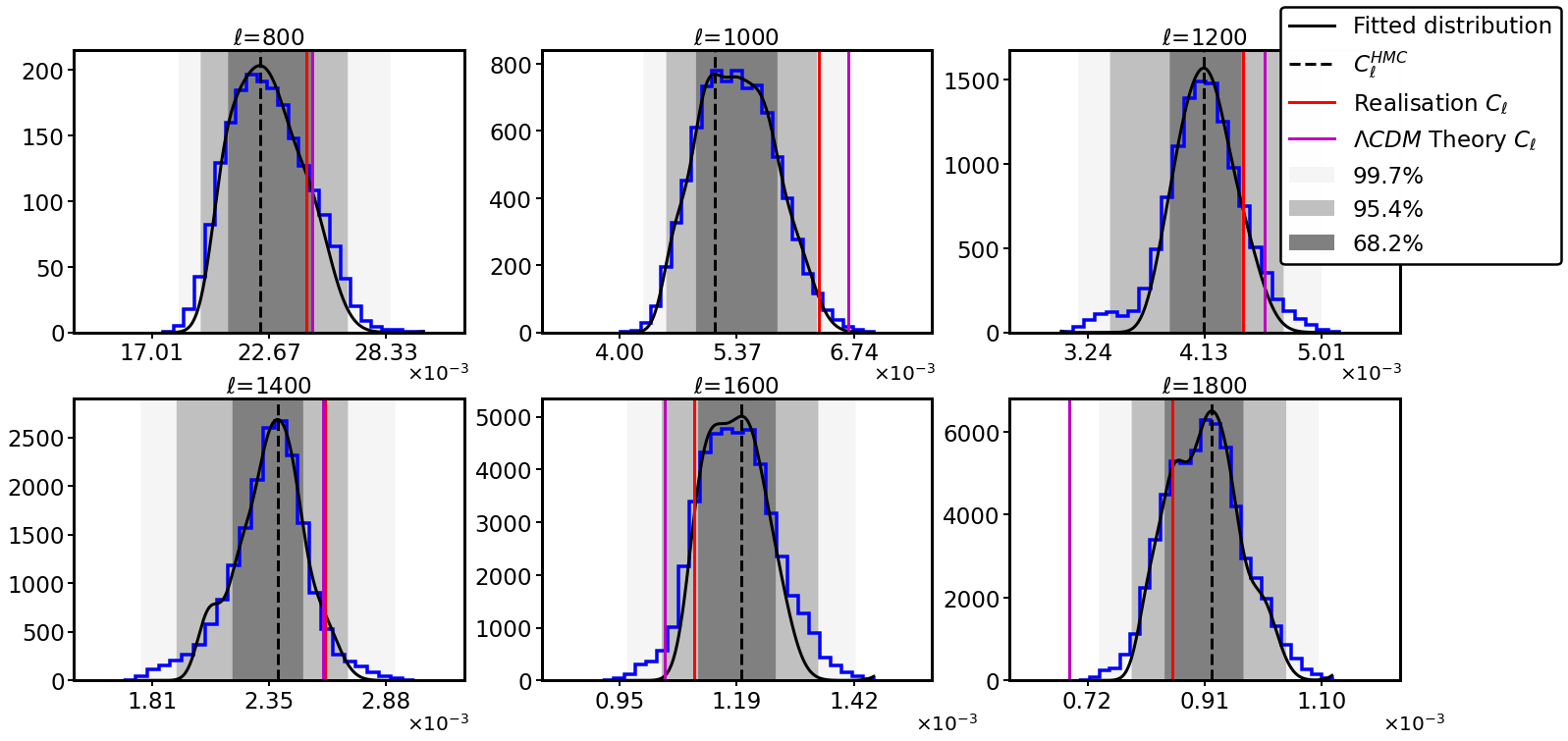}}
\subfigure[ ]{\label{cl_plot_b} 
\includegraphics[width=0.8\linewidth]{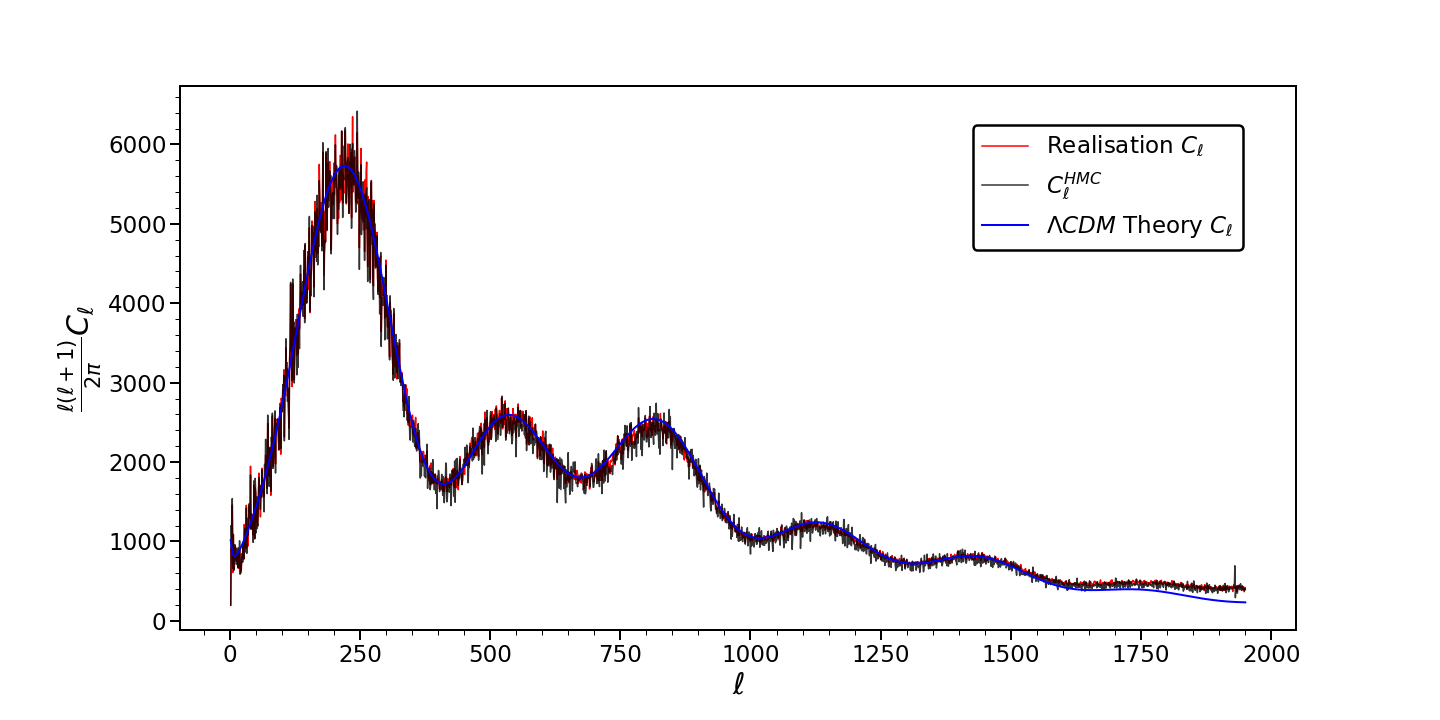}}
\caption{Top panel shows the distributions of $C_l$'s for six different $l$ values. Histograms are obtained using the samples of $C_l$. The solid black curve is the analytic distribution curve of that specific $C_l$. The black dashed line shows the peak values. The solid red line represents the value of $C_l$ obtained from \texttt{SMICA} map. The magenta line shows the best fit $\Lambda$CDM theory $C_l$ given by \textit{Planck}. Bottom panel depicts the peak values of $C_l$ distribution inferred in this analysis (black curve) along with the realisation $C_l$ (red curve) obtained from the \texttt{SMICA} map. The blue curve represents the best fit $\Lambda$CDM theory $C_l$ given by \textit{Planck}. We have an evidence in the power spectrum of a SMICA residual of an isotropic unresolved extra-galactic point source background.}
\end{figure}

 \textbf{Joint inference of the power spectrum and BipoSH spectrum}: In the rest of this section we discuss the distribution of the CMB covariance matrix parameters obtained using the HMC technique. By allowing $a_{lm}$ to vary, subject to the assumptions of the data likelihood, we assimilate all the information present in the data map. However, it is the samples of the covariance matrix parameters, $C_l$ and $\beta_{1N}$, that confront the Doppler boost model. In Figure \ref{cl_plot_a}, we present the distributions of $C_{l}$ samples for some specific values of $l$. For comparison, we also provide the realisation $C_l$ values of \texttt{SMICA} map (accounting the $N_l$ and the beam) and the $\Lambda CDM$ best-fit theory $C_l$'s given by \textit{Planck} \cite{Aghanim:2019ame}, at those $l$'s. The fitted curves in these figures are the analytic form of marginalised posterior distributions of $C_l$'s as given in \cite{Wandelt:2003uk, Larson:2006ds}
 \begin{align}\label{blackwell_rao}
     \calP (C_l|\bs{d}) \approx \frac{1}{N_{\text{sample}}}\sum_{i} \calP (C_l|\sigma_l^i),
 \end{align}
 where $\sigma_l^i$ is the realisation power spectra of the $a_{lm}$ samples and the index $i$ runs overs the $N_{\text{sample}}$ number of samples. The conditional distribution of $C_l$, given $\sigma_l^i$ has the following form \cite{Wandelt:2003uk, Larson:2006ds}
 \begin{align}
     \calP (C_l|\sigma_l^i) \propto C_l\left( \frac{\sigma_l^i}{C_l} \right)^{\left(\frac{2l+1}{2} -1\right)}\exp{\left[-\frac{2l+1}{2}\frac{\sigma_l^i}{C_l}\right]}.
 \end{align}
 We use this analytical form to fit the corresponding distributions\footnote[13]{An extremely large number of independent $\sigma_l$ samples is required in \Eq{blackwell_rao} to get the accurate distribution at high $l$. So instead, we take $N_{\text{sample}}=10$ and treat those ten $\sigma_l^i$ as free parameters in the analytical form of the distribution. We use the subroutine \texttt{scipy.optimise.curve\_fit} \cite{2020SciPy-NMeth} to optimize the values of these parameters to get the best-fit curve for the posterior distribution.}.
 We present the peaks of the distributions as the estimated value of $C_l$'s in our analysis. In Figure \ref{cl_plot_b}, we plot the inferred power spectrum along with the realisation $C_l$ of \texttt{SMICA} map and the $\Lambda CDM$ best-fit theory $C_l$'s.   
 
\begin{figure}
\includegraphics[width=0.7\linewidth,center]{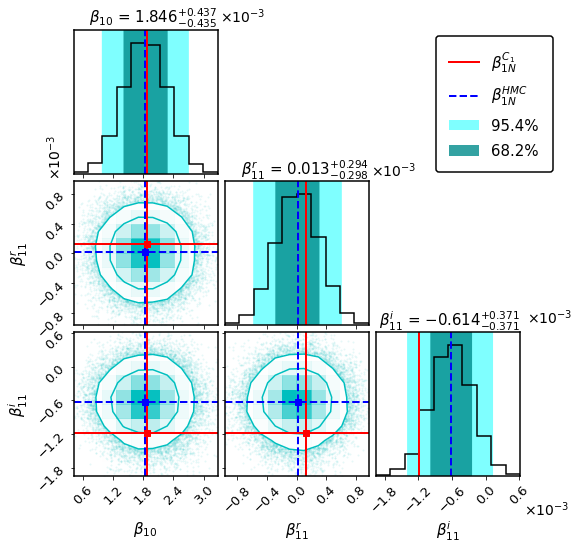}
\caption{The figure presents the 1D distributions and 2D joint distributions of $\beta_{10}$, $\beta_{11}^r$ and $\beta_{11}^i$ samples. These samples are obtained from the analysis of the \texttt{SMICA} 2018 temperature map, with $f_{sky}=40.1\%$ mask, using multipoles up to $l_{max}=1950$. The contours in the joint distribution show the levels for 68\% and 90\% of the sample points, respectively. The blue dashed line shows the peak values of the distribution and the red solid line represents the canonical value of our local motion from CMB dipole.}
\label{fig_DB_m1M}
\end{figure}

\begin{figure}
\centering
\subfigure[ ]{\label{fig_smica_A} 
\hspace*{-1.0cm}
\includegraphics[width=0.62\textwidth]{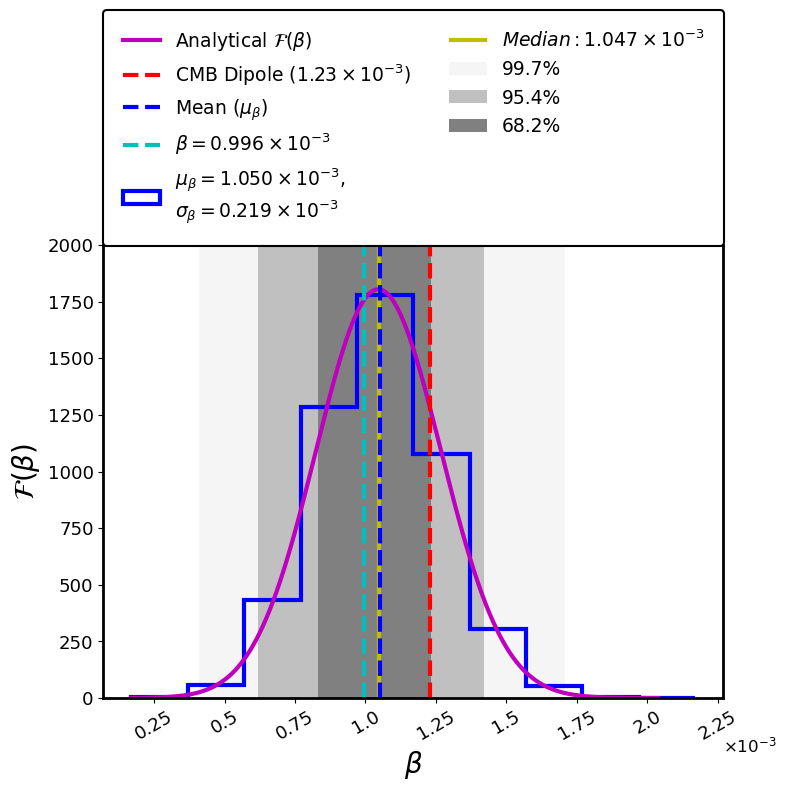}}
\subfigure[ ]{\label{fig_smica_theta_phi}
\hspace*{-0.5cm}
\includegraphics[width=0.65\textwidth]{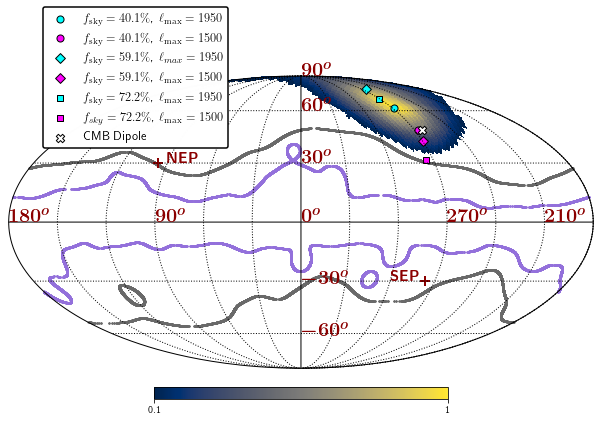}}
\caption{Figure \ref{fig_smica_A} shows the distribution of $\beta$ and Figure \ref{fig_smica_theta_phi} shows the joint distribution of $\theta_{\beta}$ and $\phi_{\beta}$ with $f_{\text{sky}}=40.1\%$ mask, using multipoles up to $l_{\text{max}}=1950$. The samples of $\beta$, $\theta_{\beta}$ and $\phi_{\beta}$ has been obtained from the $\beta_{10}$, $\beta_{11}^r$ and $\beta_{11}^i$ samples using \Eq{real_var_relation}. In Figure \ref{fig_smica_A}, the dashed blue line is the mean of this histogram, whereas the dashed cyan line depicts the $\beta$ value corresponding to the maximum posterior points of $\beta_{10}$, $\beta_{11}^r$ and $\beta_{11}^i$ parameters. The purple curve represents the analytical form of this distribution (as derived in \cite{Shaikh:2019dvb}). The dashed red line presents the known dipole amplitude corresponding to the Solar System velocity $1.23\times10^{-3}$. In Figure \ref{fig_smica_theta_phi}, the $\theta_{\beta}$ and $\phi_{\beta}$ samples are binned in a \texttt{HEALPix} grid of $N_{\text{side}}=32$, normalized by its peak value and smoothed by a Gaussian kernel. The cyan circle represents the maximum posterior point of the distribution with our primary set-up. The white cross represents the known dipole direction. In this figure, we also depict the inferred direction for the other analysis setups: $f_{\text{sky}}=59.1\%,\ l_{\text{max}}=1950$ (cyan diamond); $f_{\text{sky}}= 72.2\%,\ l_{\text{max}}=1950$ (cyan square); $f_{\text{sky}}=40.1\%,\ l_{\text{max}}=1500$ (magenta circle); $f_{\text{sky}}=59.1\%,\ l_{\text{max}}=1500$ (magenta diamond) and $f_{\text{sky}}= 72.2\%,\ l_{\text{max}}=1500$ (magenta square). The grey and purple lines show the borders of $f_{\text{sky}}=40.1\%$ and $f_{\text{sky}}=72.2\%$ masks respectively.}
\end{figure}

 Next, we present the inference of the off-diagonal terms in the covariance matrix. As the expression of the off-diagonal terms \Eq{ALM} suggests, this eventually boils down to the inference of the Doppler boost signal in our case. We have provided the summary of our results with the three choices of masking (as given in Figure \ref{fig_map1}) using multipoles up to $l_{\text{max}}=1950$, in Table \ref{tab:title1}. In this table, we present the maximum posterior points as the inferred values of the $\beta_{10}$, $\beta_{11}^r$ and $\beta_{11}^i$ parameters in our analysis, along with the error bars. We also provide the velocity amplitudes $\beta$ and the direction of the velocity in galactic coordinates ($\ell,b$), obtained from the maximum posterior points of their distributions, along with their corresponding error bars. The reported $\beta$ is obtained from the corresponding maximum posterior points of the $\beta_{10}$, $\beta_{11}^r$ and $\beta_{11}^i$ distributions. We report the signal to noise ratio (SNR) of the measurement of the Doppler boost signal for different choices of galactic mask in Table \ref{tab:title1}. A $5.23\sigma$ detection of the non-zero value of Doppler boost is achieved from the case with  $f_{\text{sky}}\,=\,72.2\%$. For the other choices of masks, $f_{\text{sky}}\,=\,59.1\%$ and $f_{\text{sky}}\,=\,40.1\%$, we have made a $4.97\sigma$ and $4.54\sigma$ detection of the Doppler boost signal respectively.
\label{fig_maps3}
 For our primary setup ($f_{\text{sky}}=40.1\%$, $l_{\text{max}}=1950$) we present the joint posterior distributions of $\beta_{10}$, $\beta_{11}^r$ and $\beta_{11}^i$ samples in Figure \ref{fig_DB_m1M}. We also compare the estimated values of Doppler boost parameters $\beta_{1N}$ in this analysis, with the canonical values known from the CMB dipole measurements $\beta= 1.23\times 10^{-3}$ \cite{1993ApJ...419....1K, Lineweaver:1996xa, 1996ApJ...473..576F, Hinshaw_2009}. The 2D joint distributions of the $\beta_{1N}$ parameters show that there is no significant correlation among them. The distribution of the velocity amplitude $\beta$ and the joint distribution of the direction coordinates ($\ell,b$) are given in Figure \ref{fig_smica_A} and Figure \ref{fig_smica_theta_phi}, respectively. The samples of the real-space variables $\beta$, $\ell$ and $b$ have been obtained from the $\beta_{10}$, $\beta_{11}^r$ and $\beta_{11}^i$ samples using the transformation relation \Eq{real_var_relation}. Galactic latitude $b$ is related to the \texttt{HEALPix} polar angle $\theta_{\beta}$ as $b = 90- \theta_{\beta}$, whereas  galactic longitude $\ell$ is the same as the \texttt{HEALPix} azimuthal angle $\phi_\beta$. We have also presented the inferred directions for three different masks with $l_{\text{max}}=1950$ and $l_{\text{max}}=1500$ in Figure \ref{fig_smica_theta_phi}. 
 
  The amplitude of the signal is consistent with the value measured from the CMB dipole, for all the three choices of mask and the values of $l_{\text{max}}$ considered in this analysis. The direction of the signal is most consistent with the choice $f_{\text{sky}}=40.1\%$ (as shown in Table \ref{tab:title1} and Figure \ref{fig_smica_theta_phi}), as it is the most conservative choice to avoid galactic contamination. We could not go beneath this sky fraction as that would mask the Doppler boost signal significantly causing loss of information in inferred amplitude. The choice of $l_{\text{max}}$ is also crucial in our analysis. As we can see from the $C_l$ inference in Figure \ref{cl_plot_b}, there is a mismatch between the best-fit theory $C_l$ given by \textit{Planck} and the realisation $C_l$ of \texttt{SMICA} at $l \geq 1500$, likely due to the unresolved point sources contributions at those scales \cite{Ade:2013kta, Aghanim:2015xee, Aghanim:2019ame}. For an isotropic distribution of point sources, we do not expect a statistically significant departure of the inferred direction of Doppler boost from the fiducial value inferred from CMB dipole \cite{1993ApJ...419....1K,1994ApJ...420..445F, Lineweaver:1996xa, 1996ApJ...473..576F, Hinshaw_2009}.  Furthermore the inferred $C_l$ from our analysis is consistent with the \texttt{SMICA} realisation $C_l$ as expected. So, we consider a galactic mask with a low available sky fraction and include CMB multipoles up to $l_{\text{max}}=1950$ to reduce the contamination from anisotropic galactic foregrounds in the estimation of Doppler boost signal, and also maximize the signal to noise ratio (SNR) of the detection of the signal, even though it costs us a slight shift in the inference of its direction (see Figure \ref{fig_smica_theta_phi}) from the fiducial value inferred from CMB dipole \cite{1993ApJ...419....1K, Lineweaver:1996xa, 1996ApJ...473..576F, Hinshaw_2009}. In future work, we will do a joint estimation of the point source and foreground contamination, along with the CMB and the Doppler boost signal to explore the reason for this minor shift in the Doppler boost direction.

\begin{table}[h]
\centering
\caption{Summary of results for different choices of mask, with $l_{\text{min}} = 800$ and $l_{\text{max}} = 1950$}
\label{tab:title1} 
\begin{tabular}{ |m{1.95cm}||m{1.95cm}|m{1.95cm}| m{1.95cm}|| m{1.95cm}| m{1.95cm}| m{1.0cm}|}
\hline
\multicolumn{7}{|c|}{$f_{\text{sky}} = 40.1\%$, SNR$=4.54$} \\
\hline
Parameter & $\beta_{10}\times10^{3}$ & $\beta^r_{11}\times10^{3}$ & $\beta^i_{11}\times10^{3}$ & $\beta\times10^{3}$ & $\ell$ & $b$ \\
\hline
Inferred Value      & 1.840   & 0.014  & -0.620  & 0.996 & $268.5^\circ$ & $61.8^\circ$\\
\hline
Standard Deviation & 0.436 & 0.298  & 0.378  & 0.219 & $49.8^\circ$ & $12.3^\circ$\\
\hline
\end{tabular}
\newline
\vspace*{1mm}
\newline
\begin{tabular}{ |m{1.95cm}||m{1.95cm}|m{1.95cm}| m{1.95cm}|| m{1.95cm}| m{1.95cm}| m{1.0cm}|}
\hline
\multicolumn{7}{|c|}{$f_{\text{sky}} = 59.1\%$, SNR$=4.97$ } \\
\hline
Parameter & $\beta_{10}\times10^{3}$ & $\beta^r_{11}\times10^{3}$ & $\beta^i_{11}\times10^{3}$ & $\beta\times10^{3}$ & $\ell$ & $b$ \\
\hline
Inferred Value      & 2.058   & 0.031  & -0.248  & 1.020 & $264.5^\circ$ & $75.2^\circ$\\
\hline
Standard Deviation & 0.427 & 0.327  & 0.366  & 0.205 & $88.2^\circ$ & $9.7^\circ$\\
\hline
\end{tabular}
\newline
\vspace*{1 mm}
\newline
\begin{tabular}{ |m{1.95cm}||m{1.95cm}|m{1.95cm}| m{1.95cm}|| m{1.95cm}| m{1.95cm}| m{1.0cm}|}
\hline
\multicolumn{7}{|c|}{$f_{\text{sky}} = 72.2\%$,  SNR=$5.23$} \\
\hline
Parameter & $\beta_{10}\times10^{3}$ & $\beta^r_{11}\times10^{3}$ & $\beta^i_{11}\times10^{3}$ & $\beta\times10^{3}$ & $\ell$ & $b$ \\
\hline
Inferred Value      & 2.104   & -0.002  & -0.509  & 1.087 & $270.4^\circ$ & $68.2^\circ$\\
\hline
Standard Deviation & 0.428 & 0.326  & 0.340  & 0.208 & $57.1^\circ$ & $10.3^\circ$\\
\hline
\end{tabular}
\newline
\newline
\end{table}

\subsection{Model Comparison with Bayes factor}\label{Bayes_factor}
Bayesian statistics provides a principled way to compare two different models. Using the conditional probabilities involved in the Bayes theorem, one can obtain the expression for the ratio of the probability of two models conditioned on a given data set. This ratio is called the Bayes factor.
It is a prime tool for the comparison of two models \cite{1939thpr.book.....J, Kass:1995loi, MOREY20166} under consideration. The expression of Bayes factor for two models $M_1$ and $M_2$ is given by
\begin{align}\label{bayes_fac_1}
    B_{M_1 - M_2} = \frac{\calP(M_1|d)}{\calP(M_2|d)}= \frac{\calP(M_1)}{\calP(M_2)}\frac{\calE(d|M_1)}{\calE(d|M_2)}.
\end{align}
In the absence of any a priori discriminating information between two models, we choose the prior probability $\calP(M)$ for the two models $M_1$ and $M_2$ to be equal. Hence the Bayes factor becomes just the ratio of the Bayesian evidence $\calE(d|M)$ for those two models. 

In our work, the two models that we compare are as follows. The $M_2$ in the denominator, is the Doppler boost model. The $M_1$ in the numerator, is the hypothesis that the CMB sky is Doppler boosted with the velocity amplitude, $\beta = 1.23\times 10^{-3}$ (the inferred velocity amplitude from CMB dipole \cite{1993ApJ...419....1K, Lineweaver:1996xa, 1996ApJ...473..576F, Hinshaw_2009}). For the sake of convenience, we reparametrize our variables to equal variance parameters \cite{Shaikh:2019dvb},
\begin{align}
    w_z = \beta_{10}, w_x = -\sqrt{2}\beta_{11}^r, w_y = \sqrt{2}\beta_{11}^i.
\end{align}
In this new three dimensional parameter space, the norm $r$ can be defined as
\begin{align}
    r \equiv \sqrt{w_z^2 + w_x^2 + w_y^2} = \sqrt{\beta_{10}^2 + 2{\beta_{11}^r}^2 + 2{\beta_{11}^i}^2} = \sqrt{\frac{4\pi}{3}}\beta.
\end{align}
So the model $M_1$ is a nested model within $M_2$, with the value of the parameter $r=\left. r \right\rvert_{\beta=1.23\times10^{-3}}$. Hence, the Bayes factor can be obtained using the Savage-Dickey density ratio (SDDR) \cite{10.1214/aoms/1177693507, Trotta:2008qt},
\begin{align}
    B_{\text{SDDR}} =\left. \frac{\calP_{C_l, \theta, \phi} (r|d, DB)}{\Pi (r|DB)}\right\rvert_{\beta =1.23\times10^{-3}},
\end{align}
where, $\calP_{C_l, \theta, \phi} (r|d, DB)$ is the posterior distribution of the norm ($r$), marginalised over power spectrum ($C_l$) and the direction ($\theta, \phi$). $\Pi$ is the prior distribution of the norm. We choose $\Pi$ to be uniform within a sphere of radius $R=\left. r \right\rvert_{\beta=2.73\times10^{-3}}$, centered at $(w_x, w_y, w_z) = (0,0,0)$, and zero outside. The marginalized prior density of the norm is
\begin{align}
    \Pi (r|DB) = \frac{3 r^2}{R^3} \ \ \ \ \text{for }r \leq R.
\end{align}
With this setup, we find the Bayes factor in favour of $M_1$, to be
\begin{align}
    B_{\text{SDDR}} \approx 7.43 \,\, \text{for $f_{sky}= 40.1\%$}.
\end{align}
 According to the Jeffreys' scale \cite{1939thpr.book.....J}, this indicates that the data substantially favors the value of $\beta$ from known CMB dipole, as it stays within $1.1\sigma$ from the inferred signal strength in our analysis. For the other two masks ($f_{\text{sky}}=59.1\%$ and $72.2\%$), the values of the Bayes factor in favor of the canonical value of $\beta$, are $8.14$ and $8.72$, respectively. The Bayes factors with $l_{\text{max}}=1500$ are $5.22$, $5.57$ and $4.99$ for $f_{\text{sky}}=40.1\%$, $59.1\%$ and $72.2\%$ respectively.

We also calculate the Bayes factor for the $\beta$ value inferred using the amplitude of the dipole signal $0.01554$ estimated in \cite{Secrest:2020has}, which makes use of the CatWISE2020 quasar catalog \cite{Eisenhardt_2020} prepared from the WISE \cite{Wright_2010} and NEOWISE data set \cite{2018RNAAS...2....1M}. By using the median value of $1.17$ for the power-law spectra index (parameter $\alpha$ in Eq. 1 in  \cite{Secrest:2020has}), we get a value of  $\beta=2.73\times10^{-3}$ from the observed dipole amplitude $0.01554$ \footnote[14]{For a mean value of the power-law spectral index $\alpha=1.26$, one can get a value of $\beta=2.66\times10^{-3}$ \cite{priv_com}.}.  
Now the $M_1$ in \Eq{bayes_fac_1} is the hypothesis that the CMB sky is Doppler boosted with an velocity amplitude $\beta = 2.73\times10^{-3}$\cite{Secrest:2020has}. We do not have enough samples at this value of $\beta$ to compute the value of posterior from the histogram. Hence, we use the analytical form of the posterior distribution to obtain the value at the desired point. Since $w_x$, $w_y$ and $w_z$ are all Gaussian random variables, we can derive the analytic form of the posterior distribution of norm ($r$), marginalized over $(\theta, \phi)$ to be \cite{Shaikh:2019dvb}
\begin{align}
    \calP (r) = \frac{r}{\sigma ^2}\sqrt{\frac{r}{r_\ast}}\exp \left[-\frac{r^2 +r_\ast^2}{2\sigma^2}\right]I_{1/2}\left(\frac{rr_{\ast}}{\sigma^2}\right) \ \ \ \ \text{for }r \leq R,
\end{align}
where $r_{\ast}$ is the norm of $(w_{x\ast},w_{y\ast}, w_{z\ast})$, the maximum likelihood point in the $w_{x}w_{y}w_{z}$-space and $\sigma$ is the standard deviation of each of these variables. $I_{1/2}$ is the modified Bessel function of first kind with order $1/2$. We estimate the values of $r_{\ast}$ and $\sigma$ from the $(w_x, w_y, w_z)$ samples. Using the value of posterior and prior at $\beta = 2.73 \times 10^{-3}$, we find the Bayes factor in favour of this model to be
\begin{align}
    B_{\text{SDDR}} \approx 1.24\times10^{-11}.
\end{align}
This indicates a strong disagreement (according to the Jeffreys' scale \cite{1939thpr.book.....J}) with the value of local motion $\beta= 2.73\times 10^{-3}$. The mismatch between the amplitude of the velocity inferred from CMB and from quasars differs by about $8 \sigma$. For the other two choices of masks, ($f_{\text{sky}}=59.1\%$ and $72.2\%$), the values of the Bayes factor in favour of the quasar value are $3.35\times10^{-11}$ and $6.58\times10^{-11}$, respectively, indicating strong disagreement with the value inferred in this work. The Bayes factors with $l_{\text{max}}=1500$ are $1.47\times10^{-5}$, $1.65\times10^{-4}$ and $2.26\times10^{-4}$ for $f_{\text{sky}}=40.1\%$, $59.1\%$ and $72.2\%$ respectively. So, for both the extreme choices of $l_{\text{max}}$ considered in this analysis, a high value of local motion ($\beta \approx 2.73 \times 10^{-3}$) is strongly disfavoured\footnote[15]{We also calculate the Bayes factor for $\beta = 2.66 \times 10^{-3}$ which corresponds to $\alpha_{\text{mean}}=1.26$ of the distribution given in \cite{Secrest:2020has}. These values are $1.17\times10^{-10}$, $2.90\times10^{-10}$ and $5.96\times10^{-10}$ with $l_{\text{max}}=1950$ for $f_{\text{sky}}=40.1\%$, $59.1\%$ and $72.2\%$ respectively.}.

\section{Conclusions}\label{Conclusion}
In this work, we study the violation of statistical isotropy of the CMB sky due to the Doppler boost of CMB photons in our observation frame. We have assumed the statistical isotropy of CMB fluctuations in the cosmological rest-frame. In this joint inference, we explore the parameter space of the whole CMB covariance matrix under the Doppler boost model in a Bayesian framework. This approach is necessary because the anisotropic off-diagonal part is influenced by the isotropic diagonal part. The inference of the off-diagonal elements leads to the estimation of $\beta_{1N}$ parameters (the amplitude and direction of the local motion) as in \Eq{biposh_1}. We jointly infer our model parameters, i.e. the spherical harmonic coefficients $a_{lm}$'s, the power spectra $C_l$'s and the Doppler boost signal $\beta_{10}$, $\beta_{11}^r$ \& $\beta_{11}^i$. This is achieved by sampling the posterior distribution \Eq{eq_post_s_CS} for \textit{Planck}-2018 \texttt{SMICA} temperature anisotropy map. Inference of the high-dimensional parameter space is facilitated by the use of the HMC sampling method. The frequency dependence of the modulation effect in the \texttt{SMICA} temperature map is captured using a \texttt{SMICA} boost factor $b^{\texttt{SMICA}}(l)$, which is a weighted sum of the frequency dependence with the SMICA weights. The expression of {$b^{\texttt{SMICA}}(l)$} is given in \Eq{b_eff_nu_lm} and plotted in Figure \ref{eff_bnu} for different choices of mask. This recipe can also be used for any CMB map estimate obtained using the internal linear combination based method and can be used in relevant future work.

We have performed our analysis on three different sky fractions ($f_{\text{sky}} = 40.1\%, 59.1\%\text{ and } 72.2\%$) using multipoles from $l_{\text{min}}=800$ up to $l_{\text{max}}=1950$. With these setups, we have detected the non-zero value of the Doppler boost signal which is consistent with the fiducial values from the known CMB dipole \cite{1993ApJ...419....1K,1994ApJ...420..445F, Lineweaver:1996xa, 1996ApJ...473..576F, Hinshaw_2009}, with a significance of $4.54\sigma$, $4.97\sigma$ and $5.23\sigma$, respectively, as expected from a previous forecast \cite{Mukherjee:2013zbi}. With our primary setup of $f_{\text{sky}}=40.1\%$ and $l_{\text{max}}=1950$, we conclude that our analysis finds value of the peculiar velocity of our observation frame with respect to the CMB to be $v = (298.5\pm65.6)$ km/s. The direction of this velocity has been found in galactic coordinates $(\ell,b)$ = ($268.5^\circ\pm49.8^\circ$, $61.8^\circ\pm12.3^\circ$). We have computed the Bayes factor in favour of the canonical values from the known CMB dipole, to be $7.43$. This value indicates a definite evidence of the canonical values in CMB data. The Bayes factor in favour of the dipole value inferred from the quasar distribution \cite{Secrest:2020has} is $\sim 1.24 \times 10^{-11}$, indicating strong disagreement according to the Jeffreys' scale \cite{1939thpr.book.....J}. Similar analysis has also been carried out using \texttt{NILC} and \texttt{SEVEM}, with reasonable approximations for the effective boost factor. The values of inferred velocity and direction\footnote[16]{However, we don't put the detailed results for \texttt{NILC} and \texttt{SEVEM} due to the unavailability of their respective weights and the corresponding propagation codes in Planck Legacy Archive and IRSA websites. These are required to calculate the effective boost factor, which is essential to estimate the Doppler boost signal.} have been found to be consistent with the ones from \texttt{SMICA}. The consistency of the Doppler boost signal presented in this work method from the small angular scale temperature data with the measured CMB dipole suggests evidence for the absence of other dipole modulation signal with amplitude higher than $\sim 10^{-3}$ at those angular scales. This is consistent with the power-law decay of the cosmic hemispherical asymmetry signal in $l$ obtained previously \cite{Shaikh:2019dvb}; and with the non-detection of directional anomalies in cosmological parameter estimates from the small scale anisotropies \cite{Mukherjee:2015mma,Mukherjee:2017yxd}. 

In our analysis, we have used a diagonal noise covariance matrix, computed from 200 FFP10 noise simulations. However a further improvement can be made by also including the off-diagonal terms of the covariance matrix. This will be explored in a future work using a fast covariance matrix calculation technique \cite{Ramanah:2019rpg}. This Bayesian formalism can also be applied to the \textit{Planck} polarization map to extract the Doppler boost signal  \cite{Mukherjee:2013zbi} using the \texttt{SMICA} boost factor introduced in this analysis. The method will also be useful for the future ground-based and space-based high resolution CMB experiments such as Advanced ACTPol \cite{Henderson:2015nzj}, SPT-3G \cite{Benson:2014qhw}, Simons Observatory \cite{Ade:2018sbj},  CMB-S4 \cite{Abazajian:2019eic}, CMB-HD \cite{Sehgal:2020yja}, CMB-Bh\=arat\footnote[17]{\url{http://cmb-bharat.in}}, and PICO \cite{Hanany:2019lle}  to infer the value of local motion \cite{Challinor:2002zh,Amendola:2010ty,Mukherjee:2013zbi} and also the imprints of cosmic hemispherical asymmetry \cite{Mukherjee:2014lea, Mukherjee:2015wra,Cayuso:2019hen} from CMB temperature and polarization anisotropy.

\vskip20pt
\textbf{Acknowledgements:} S. Mukherjee thanks Jean Francois Cardoso for providing the SMICA weights. S. Mukherjee also acknowledges useful discussions with Sebastian von Hausegger, Roya Mohayee, Mohammed Rameez, and Subir Sarkar. The authors also thank Subir Sarkar and Jens Chluba for their valuable comments on the paper. S. Saha acknowledges Institute fellowship from IISER-Pune, MHRD. S. Shaikh acknowledges Science and Engineering Research Board of the Department of Science and Technology, Govt. of India, grant number SERB/ECR/2018/000826 for funding the position at NISER, India. The work of S. Mukherjee is supported by the Delta ITP consortium, a program of the Netherlands Organisation for Scientific Research (NWO) that is funded by the Dutch Ministry of Education, Culture, and Science (OCW). Research at Perimeter Institute is supported in part by the Government of Canada through the Department of Innovation, Science and Economic Development Canada and by the Province of Ontario through the Ministry of Colleges and Universities. The work of BDW is supported by the Labex ILP (reference ANR-10-LABX-63) part of the Idex SUPER,  received financial state aid managed by the Agence Nationale de la Recherche, as part of the programme Investissements d'avenir under the reference ANR-11-IDEX-0004-02.  BDW acknowledges financial support from the ANR BIG4 project, under reference ANR-16-CE23-0002. The Center for Computational Astrophysics is supported by the Simons Foundation. The support and the resources provided by ‘PARAM Brahma Facility’ under the National Supercomputing Mission, Government of India at the Indian Institute of Science Education and Research, Pune are gratefully acknowledged. We acknowledge the help of \texttt{SMICA} propagation code\footnote[18]{\texttt{smica\_coadd.py} in \texttt{COM\_Code\_SMICA-weights-propagation\_R3.00.tar.gz}, available in \url{https://wiki.cosmos.esa.int/planck-legacy-archive/index.php/SMICA_propagation_code}.} for computing the effective boost factors $b_{\texttt{SMICA}}(l)$. We acknowledge the use of the following packages in this analysis: \texttt{CoNIGS}~\cite{Mukherjee:2013kga}, \texttt{Corner}~\cite{corner},  \texttt{HEALPix}~\cite{Gorski:2004by}, \texttt{healpy}~\cite{Zonca2019}, \texttt{Ipython}~\cite{4160251}, \texttt{Matplotlib}~\cite{4160265}, \texttt{NumPy}~\cite{2011CSE....13b..22V}, and  \texttt{SciPy}~\cite{2020SciPy-NMeth}.


\label{Bibliography}

\bibliographystyle{JHEP}
\bibliography{references}
\end{document}